\documentclass[manuscript]{aastex}
\shorttitle{The binary fraction of NGC 1818}
\shortauthors{Hu et al.}

\begin{document}

\title{The binary fraction of the young cluster NGC 1818 in the Large
Magellanic Cloud}

\author{Y. Hu,\altaffilmark{1,2} L. Deng,\altaffilmark{1} Richard de
Grijs,\altaffilmark{3,4,1} Q. Liu,\altaffilmark{1,2} and Simon
P. Goodwin\altaffilmark{4}}

\email{licai@bao.ac.cn}

\altaffiltext{1}{Key Laboratory of Optical Astronomy, National Astronomical
Observatories, Chinese Academy of Sciences, Beijing 100012, P. R. China}
\altaffiltext{2}{Graduate University of the Chinese Academy of Sciences, Beijing
100012, P. R. China}
\altaffiltext{3}{Kavli Institute for Astronomy and Astrophysics,
Peking University, Beijing 100871, P. R. China}
\altaffiltext{4}{Department of Physics \& Astronomy, The University of
Sheffield, Sheffield S3 7RH, UK}

\begin{abstract}
We use high-resolution {\sl Hubble Space Telescope} imaging
observations of the young ($\sim 15-25$ Myr-old) star cluster NGC 1818
in the Large Magellanic Cloud to derive an estimate for the binary
fraction of F stars ($1.3 < M_\star/M_\odot < 1.6$). This study
provides the strongest constraints yet on the binary fraction in a
young star cluster in a low-metallicity environment ($\mbox{[Fe/H]}
\sim -0.4$ dex). Employing artificial-star tests, we develop a simple
method that can efficiently measure the probabilities of stellar
blends and superpositions from the observed stellar catalog. We create
synthetic color-magnitude diagrams matching the fundamental parameters
of NGC 1818, with different binary fractions and mass-ratio
distributions. We find that this method is sensitive to binaries with
mass ratios, $q \ga 0.4$. For binaries with F-star primaries and mass
ratios $q > 0.4$, the binary fraction is $\sim 0.35$. This suggests a
total binary fraction for F stars of 0.55 to unity, depending on
assumptions about the form of the mass-ratio distribution at low $q$.
\end{abstract}


 \section{Introduction}

The majority of stars are thought to form in binary or
multiple\footnote{For brevity, by `binaries' we generally also mean
`multiples', since many stars are found in triple and higher-order
multiple systems. We will distinguish the two only when it is
important to do so. Note that the analysis in this paper is of {\em
binary} systems.} systems (Goodwin \& Kroupa 2005; Duch\^ene et
al. 2007; Goodwin et al. 2007), and the initial binary properties of
stars place important constraints on star formation and the origin of
the stellar initial mass function (IMF; Goodwin et al. 2007,
2008). The majority of stars with masses greater than approximately
$0.5 M_\odot$ are also thought to form in star clusters (Lada \& Lada
2003), and the binary content of a star cluster plays an important
role in both its observational properties and its dynamical evolution
(e.g., Kroupa et al. 1999). In addition, many exotic objects observed
in star clusters, such as blue stragglers, cataclysmic variables, and
X-ray sources, are believed to be related to binary systems. Almost
all studies of binarity have been limited to nearby, solar-metallicity
populations. However, it might be expected that metallicity (e.g.,
through its effects on cooling and, hence, on the opacity limit for
fragmentation) will play a role in the fragmentation of cores to
produce binary systems (Bate 2005; Goodwin et al. 2007).

In general, the most direct way in which to study binary fractions is
by examining whether a given star is part of a binary system, on an
individual basis. Over the past two decades, the binary fractions of
field stars in the solar neighborhood have been studied carefully in
this conventional fashion (e.g., Abt 1983 for B stars; Duquennoy \&
Mayor 1991 for G dwarfs; Fischer \& Marcy 1992 for M dwarfs;
Kouwenhoven et al. 2006 for A and B stars; see also Goodwin et
al. 2007 for a review). Nearby clusters and associations have also
been examined in detail (see Duch\^ene 1999 and Duch\^ene et al. 2007
for reviews and comparisons). However, the binary fractions in more
distant, massive clusters have not yet been studied thoroughly,
because these environments are too crowded and their distances too
great, so that their member stars are too faint to be examined
individually for binarity. Fortunately, there is an alternative
approach, i.e., by means of an artificial-star-test technique, which
allows us to estimate the binary fractions in crowded environments. By
studying the morphology of their color-magnitude diagrams (CMDs),
Romani \& Weinberg (1991) determined the observed binary fractions in
M92 and M30 at $\la 9$ and $\la 4$\%, respectively, for the full
populations of cluster stars down to $m_V \sim 23$ and $\sim 22$ mag,
respectively, based on two-dimensional maximum-likelihood
analysis. Rubenstein \& Bailyn (1997, hereafter RB97) investigated the
binary fraction of main-sequence stars with $15.8 < V < 28.4$ mag in
the $\sim 13.5$ Gyr-old (e.g., Pasquini et al. 2007)
post-core-collapse Galactic globular cluster NGC~6752. They found a
binary fraction of 15--38\% inside the inner core radius, falling to
$\la 16$\% at larger radii, with a power-law mass-ratio
distribution. For other old globular clusters, Bellazzini et
al. (2002) estimated the binary fraction in NGC 288 for stars with $20
< V < 23$ mag (corresponding to masses of $0.54 \la M_\star/M_\odot
\la 0.77$) at 8--38\% inside the cluster's half-mass radius (and at $<
0.10$ in the outer regions, most likely close to zero), regardless of
the actual mass-ratio distribution. Zhao \& Bailyn (2005) claimed
6--22\% of main-sequence binaries ($19.2 \le m_{\rm F555W} \le 21.2$
mag) for M3, within the cluster's core radius, and between 1 and 3\%
for stars between 1 and 2 core radii. By applying similar techniques
to the post-core-collapse Galactic globular cluster NGC 6397, Cool \&
Bolton (2002) derived a binary fraction of 3\% for main-sequence stars
with primary masses between 0.45 and $0.80 M_\odot$, for a range of
mass ratios. Based on an extrapolation to all mass ratios, they
estimated the total main-sequence binary fraction in the cluster core
at 5 to 7\%. Davis et al. (2008), using the method pioneered by Romani
\& Weinberg (1991) in combination with numerical simulations by Hurley
et al. (2007), concluded that the outer regions of this cluster (at
1.3--2.8 half-mass radii) exhibit a binary fraction ($1.2 \pm 0.4$\%)
close to the primordial fraction of $\sim$1\% predicted by the
simulations, while the inner region has a substantially higher
fraction, $5.1 \pm 1.0$\%. However, {\it all clusters thus far studied
in this way are old stellar systems} (cf. table 1 in Davis et
al. 2008) in which dynamical evolution is expected to have
significantly altered the initial binary population.

In this paper, we use accurate photometric observations of the young,
low-metallicity star cluster NGC 1818 in the Large Magellanic Cloud,
taken with the Wide Field and Planetary Camera-2 (WFPC2) on board the
{\sl Hubble Space Telescope (HST)}, to study its binary fraction. The
photometric data and the cluster CMD are discussed in \S 2. Our newly
developed method to correct for stellar blends and superpositions,
based on the artificial-star-test technique, is presented in \S 3. The
fitting of the binary fraction is discussed in \S 4, and \S 5 contains
a further discussion and our conclusions.

\section{Observations, data reduction, and background decontamination}

Our data set was obtained from {\sl HST} program GO-7307 (PI Gilmore),
which included three images in both the F555W and F814W filters (with
exposure times of 800, 800, and 900s per filter). Specifically,
we used the observations with the PC1 chip centered on the cluster's
half-light radius (see Fig. 1). The origin of this
data set and the program's scientic rationale were described in detail
in de Grijs et al. (2002a, and references therein). The observations
were reduced with {\sc HSTPhot} (version 1.1, May 2003; Dolphin 2002)
using the point-spread-function (PSF)-fitting option. Bad and (close
to) saturated pixels were masked using data-quality images. Cosmic
rays were removed using {\sc HSThphot}'s {\sc crmask} routine. The
three images were combined into a single deep frame (with a total
exposure time of 2500s) using the {\sc coadd} routine and hot pixels
were removed following the procedure recommended in the {\sc HSTphot}
manual. Photometry was performed on both the F555W and F814W images
simultaneously, using a weighted PSF fit (option `2048' in {\sc
HSTphot}), as suggested by Holtzman et al. (2006). The instrumental
magnitudes were converted to the standard Johnson/Cousins $V$ and $I$
filters using the transformation solutions of Dolphin (2000).

We show the CMD around the cluster's half-light radius (as
defined in HST program GO-7307) in Fig. 2 (left-hand
panel) and provide the current-best estimates of a few important
cluster parameters including the core and half-mass radii, in
Table 1, as well as the relevant bibliographic references. We use the
Padova isochrones (Girardi et al. 2000) to perform our fits to the
cluster's CMD (see Fig. 3, left-hand panel). For
comparison, we also show the CMD obtained by Liu et al. (2009; see
also de Grijs et al. 2002a) using the {\sc iraf apphot} (aperture
photometry) software package in Fig. 2 (right-hand
panel). Our newly determined CMD is cleaner and the main sequence is
much tighter than that of Liu et al.'s (2009) CMD, because the HSTphot
software package we used is much better at properly handling stellar
photometry in crowded fields than {\sc iraf}'s {\sc apphot} routine, while
 our PSF-fitting approach ensures higher-precision
photometric measurements than Liu et al.'s (2009) aperture photometry
(cf. Fig. 2). We will provide a careful, quantitative comparison of
the results from both approaches in Hu et al. (in prep.).

As noted by Castro et al. (2001), an old red-giant population and an
intermediate-age red-giant clump are clearly seen in the CMD of NGC
1818. If we adopt an age for this cluster of approximately 25 Myr
(e.g., de Grijs et al. 2002a; and references therein), these older
components can only be interpreted as background field stars in the
LMC's disk. Therefore, the main sequence of NGC 1818 is severely
contaminated by field stars.  Here, we adopt a statistical approach
similar to that adopted by Bonatto et al. (2006) to subtract
background stars.  The background data set (to which we applied
exactly the same photometric procedures as for our science field) was
obtained as part of {\sl HST} program GO-6277 (PI Westphal); it is
suitably located at a distance of $\sim$8 arcmin from the cluster's
half-mass radius.\footnote{We initially selected a number of suitable
LMC fields for our background substraction, including the field
specifically associated with the cluster from {\sl HST} program
GO-7307. Eventually, we decided to use the field that best represented
the background isochrone shown in the left-hand panel of Fig. 2.}
 The middle panel of Fig. 3 shows the CMD of the LMC
background field near NGC 1818, which was specifically observed for
background characterization (see, for more details, Castro et
al. 2001; Santiago et al. 2001; de Grijs et al. 2002a). We only remove
the stars in the region $19\leq V \leq25$ mag and $-0.1\leq (V-I) \leq
1.5$ mag, since this region contains almost all stars for which the
observational completeness fraction is greater than 50\%. To perform
the field-star decontamination procedure, we divide both the
background and cluster CMDs into grids of cells, in color and
magnitude. We count the number of stars in each cell in the background
CMD, and then randomly remove the corresponding number of stars,
corrected for the difference in area covered, from the respective cell
of the cluster CMD. The choice of cell size affects the appearance of
the resulting background-subtracted cluster CMD significantly. After
extensive experimentation we chose a cell size of three times the
observational uncertainty in both magnitude and color of single
stars (i.e., $3 \sigma = 0.06$ mag) to minimize any significant
effects due to stochasticity. The right-hand panel of
Fig. 3 shows the results of the background
decontamination.

We performed additional tests to validate our approach, based on the
simplest assumption that if we subtract the background stellar
population from the original background field, we should be left with
only statistical noise, which should, therefore, not lead to
systematics in our analysis. These tests indeed showed that our
background-subtraction procedure is robust. Nevertheless, a close
examination of the right-hand panel of Fig. 3 shows that
there is some residual contamination from the background stellar
population, as indicated by the presence of a faint red-giant
branch/clump feature. This is most likely caused by the local
background in the cluster region being somewhat different from that in
our comparison field (we applied two slightly different field regions
in an attempt to optimally subtract the local cluster background, but
a small residual effect remains). For the statistical comparison
carried out in the remainder of this paper, we are confident that this
residual background population affects our results negligibly,
however. We base this conclusion on (i) the fact that the dominance of
the background population is significantly redward with respect to the
expected locus of NGC 1818's main-sequence binary population (i.e.,
the residual background contamination  dominates the region {\it
outside} the parallellogram shown in the right-hand panel of
Fig. 3; our background-subtraction procedure at and near
the single- and binary-star main sequence is unaffected by these
stellar types) and (ii) a statistical analysis of the relative
importance of the genuine binaries versus the residual background
contamination: we estimate that in the region in the CMD space of
interest (i.e., the parallellogram), the fractional contamination by
the background population (i.e., our systematic error) is $\la 3$\%
(based on a detailed examination of the stellar population using star
counts). Finally, in Fig. 4 we present the completeness
curves for both the NGC 1818 observations and the background field,
for two of the four WFPC2 chips. We emphasize that we performed
completeness analyses for all four chips, so that we can properly and
consistently correct our observational data for the effects of
incompleteness. We also note that for the entire magnitude range of
interest, $V \la 22$ mag (see the parallellogram in
Fig. 3), the completeness of our data is well in excess
of 80\%, so that corrections for incompleteness are straightforward.

\section{Artificial-star tests}

Ideally, if there is no binary population, nor any observational
errors, all stars in a cluster should lie on the same isochrone,
because they were all born at approximately the same time in the same
giant molecular cloud (i.e., they have the same metallicity). However,
in Fig. 3 we clearly see a broadening of the cluster's
main sequence. There are three factors that contribute to this
broadening: (i) photometric errors, (ii) superposition effects, and
(iii) the presence of true binary and/or multiple systems.
Photometric errors broaden the main sequence symmetrically if we
assume the magnitude errors to be Gaussian, which is not unreasonable
(given that the magnitude range of interest is well away from the
observational completeness limit). However, the other two factors skew
the stars to the brighter, and redder, side relative to the
corresponding best-fitting isochrone. However, it is difficult to
distinguish between superpositions and physical binaries on the basis
of only CMD morphological analysis. To obtain the binary fraction of
NGC 1818, we therefore perform Monte Carlo tests, where we produce
artificial-star catalogs and compare the spread of real and artificial
stars around the best-fitting isochrone.

Since the photometric errors of the observed stars strongly depend on
their magnitudes and their positions on the {\sl HST}/WFPC2 chips used
for the observations, exponential functions (numerically following the
densest concentration of data points, closely matched to their lower
boundary) were adopted to fit the relation between the magnitude and
standard deviation of the photometric errors (these relations vary
between the WFPC2 chips; we have taken great care to use the
appropriate relations for our analysis). Each artificial star (see
below) is randomly assigned Gaussian photometric errors, of which the
standard Fig. 5 shows the photometric uncertainties
as a function of $V$ and $I$ magnitude for the {\sl HST}/WF3
observations (containing 2473 stars); the center of the
corresponding PC1 chip (containing 886 stars) is located at the
half-light radius. The solid curves in Fig. 5 show the
functions adopted to assign uncertainties to the individual
artificial stellar magnitudes; 80\% of the data points are located
within $\sim$1.2 times the standard deviation from the curve.

The global stellar mass function of NGC 1818 is well approximated by a
Salpeter (1955) power law for masses $>0.6 M_\odot$ (de Grijs et
al. 2002b; Kerber et al. 2007; Liu et al. 2009).\footnote{We note that
these mass-function fits were obtained on the basis of a single-star
mass-function assumption. However, as Kerber et al. (2007; see also
Liu et al. 2009) illustrate in their Fig. 9, the inclusion of binaries
affects the derived mass-function slope only slightly: the deviation
between the input (single-star) and output (single+binary) mass
functions they derive is less than 0.15 (in units of the mass-function
slope, $\alpha$), as long as the input slope is sufficiently close to
the Salpeter (1955) index and assuming a 100\% binary fraction. A
smaller binary fraction, as found in this study, will change the
mass-function slope proportionately less (see Kerber et al. 2007).}
Therefore, we draw the masses of single stars and the primary stars of
the binary population from a Salpeter (1955) $\alpha=2.35$ power-law
IMF in the mass range $0.6 \leq M_{\star}/M_\odot \leq 6.0$. The
masses of the secondaries are drawn from a given mass-ratio
distribution for all primary masses (we discuss the choice of the
mass-ratio distribution in detail in the next section). We note that
this produces a {\em total} IMF (of single stars plus each component
of binary systems) which is {\em not} equal to a Salpeter
IMF. However, the deviation from a Salpeter IMF is fairly minor (see
Kerber et al. 2007). The alternative is to draw both primary and
secondary masses from a Salpeter IMF. However, random pairing is
excluded in all observed multiple populations (see, e.g., Duquennoy \&
Mayor 1991; Fischer \& Marcey 1992; Kouwenhoven et al. 2005; Duch\^ene
et al. 2007).

The magnitudes and colors of all artificial stars are then obtained by
interpolating from the relevant isochrone. For binary stars, we simply
add the fluxes of the primaries and secondaries to obtain the
magnitudes and colors of the system. The results from this procedure
are shown in Fig. 6 (left-hand panel).

The best way to simulate the superposition effect is by adding
artificial stars to the original images (see details in RB97; and
references therein). Alternatively, in each run we randomly distribute
$5 \times 10^6$ artificial stars on the spatial-distribution diagram
of the real stars (while properly accounting for the
position-dependent photometric uncertainties using the chip-dependent
magnitude--uncertainty relations discussed above), instead of on the
original images. If an artificial star has an angular distance from
any real star within 2 pixels (corresponding to the size of our
aperture), it is assumed to be `blended' [see below; see also Reipurth
et al. (2007) for a blending analysis relevant to this study]. Its new
magnitude and color are re-calculated in the same way as for a binary
system. To avoid double counting, if the output $V$-band magnitude of
any artificial star is 0.752 mag brighter than its input magnitude (as
expected for equal-mass binary systems), we assume that we are dealing
with a chance superposition and remove the star from the output
catalog. However, we do not allow the artificial stars to blend with
each other, even when their angular distance is less than 2
pixels. Therefore, we do not need to add artificial stars multiple
times to avoid blends between them. This is one of the main
differences between our novel approach and that followed by RB97 (the
latter authors added much smaller numbers of artificial stars to their
images, precisely to avoid this blending issue).\footnote{Using this
approach, we can generate any number of artificial stars, with which
we can construct an artificial-star catalog that has the same
luminosity function and spatial distribution as the observations, thus
requiring much less computing time than when using the full images for
our analysis.} The CMDs of the artificial stars are shown in
Fig. 6. Fig. 8 presents a comparison of the
observed, background-subtracted CMD with the artificial CMD containing
50\% binaries. By adding artificial stars to the raw images, we
robustly verified that the 2-pixel threshold we adopted is a good
approximation to simulate stellar blending (see Hu et al., in prep.,
for the full quantitative analysis).

For each observed star, we find all artificial stars in the input
catalog that are located within 20 pixels and within 0.2 mag in
brightness.\footnote{This choice is driven by the need to have a
statistically sound procedure: we generated artificial stars that may
or may not fall exactly on top of a real star. Therefore, we needed to
choose a region around any real star in which any artificial star(s)
correspond(s) to the real star for statistical comparison
purposes. RB97 used stars within 100 pixels, but since we are less
restricted by computational power, a smaller radial range ensures that
the spatial distributions of the real stars and the final catalog of
artificial stars are statistically the same.} We randomly extract one
of these artificial stars as the counterpart to the observed
star. Finally, we construct a synthetic catalog containing the same
total number of {\em systems} (be they single stars or unresolved
binary systems), a similar luminosity function, projected surface
number density, and superposition probability as the original
data. The only differences between the observed and synthetic catalogs
are the binary properties (both the binary fraction and the mass-ratio
distribution).

\section{The binary fraction of NGC~1818}

We analyze stars in the mass range from 1.3 to $1.6 M_\odot$ (roughly
F stars), in the region of the CMD in the parallelogram shown in the
right-hand panel of Fig.~3, which is where any binary
sequence will show optimally. For brighter stars, the isochrone is
almost vertical, and therefore the binary sequence is too close to the
main sequence to allow us to distinguish it. (In addition, we are not
fully convinced that all these bright stars are really main-sequence
stars rather than blue stragglers.) For fainter stars, the larger
photometric errors and lower completeness make it difficult to detect
the binary population. (As we noted above, we applied
position-dependent completeness corrections to our data, which were
all $>80$\% complete in the magnitude range of interest.) Since the
isochrone is almost linear in the region in CMD space of interest, we
rotate the ordinates of the artificial and observed CMDs such that the
isochrone becomes vertical to a new `pseudo-color,' i.e., a new
function of $V$ and $V-I$ produced by rotating the CMD
(corresponding to the color axis projected onto the long side of the
parallellogram in Fig. 2). Note that the exact form of the
pseudo-color function is unimportant.

In Fig.~9 we show the cumulative distribution function
(CDF) with pseudo-color of the true CMD (solid line) and a stellar
population with photometric errors, but {\em no} binaries (dotted
line).\footnote{As the luminosity function of the final output
artificial stars is similar to the observed luminosity function, the
precise form of the stellar mass function is unimportant in this
context.} This is clearly a very poor fit to the data. We also show
the best-fitting binary fraction ($f_{\rm b}$), with a uniform $q$
distribution, of $f_{\rm b} = 0.55 \pm 0.05$ ($1\sigma$) (dashed
line), as well as the relevant model for 100\% binarity (long-dashed
line).

It is expected that this method will be insensitive to low-$q$ systems
in which the secondary component contributes very little to the
pseudo-color. To test our sensitivity to the mass ratio of a system,
we produce artificial catalogs that do not include binaries below some
mass ratio $q_{\rm cut-off}$, but contain the same number of binaries
above $q_{\rm cut-off}$. Therefore, we need to compare these results
based on the artificial-star catalogs with the best-fitting $f_{\rm b}
= 0.55$ with $q_{\rm cut-off} = 0$ (see Fig.~9).

We find that there is very little difference in the fits to the CDF
for $q_{\rm cut-off} < 0.4$, showing that we are insensitive to
binaries with mass ratios smaller than $q \sim 0.4$. In
Fig.~9 we plot the binary fraction versus $\chi^2$
probability (rms error) for different $q_{\rm cut-off}$'s. For $q_{\rm
cut-off} < 0.4$, the best fits of binary fraction are acceptable,
since the maximum probabilities are much larger than the commonly
adopted limit of $\Delta \chi^2 = 0.05$.\footnote{G. E. Dallal (2007),
http://www.jerrydallal.com/LHSP/p05.htm.} However, for $q_{\rm
cut-off} =0.5$, the value is just a little greater than 0.02. And for
$q_{\rm cut-off} = 0.7$, the very low $\chi^2$ value indicates that
even the best fit is poor. Therefore, we adopt a conservative limit to
the mass ratios to which we are sensitive of $q > 0.4$.

For this discussion, we have assumed that the $q$ distribution is flat
(at least above $q=0.4$). This is consistent with the mass-ratio
distribution of A- and B-type stars in Sco OB2 (Kouwenhoven et
al. 2005, their fig.~14). However, G-dwarf mass ratios in the solar
neighborhood are concentrated toward low $q$ (Duquennoy \& Mayor
1991). Duquennoy \& Mayor (1991) show that the mean local G-dwarf $q$
distribution shows a roughly linear decrease with increasing $q$ for
$q>0.4$ (their fig.~10).

The results indicate that the {\em total} binary fraction of F stars
in NGC~1818 is $\sim 0.55$, with an approximately flat mass-ratio
distribution. However, since we are only sensitive to binaries with $q
> 0.4$, we may only constrain the binary fraction in this mass range
to $f_{{\rm b} (q>0.4)} \sim 0.35$. It is impossible to determine the
total binary fraction without making some assumptions about the form
of the $q$ distribution below 0.4. It is unlikely that the mass-ratio
distribution declines below $0.4$, meaning that a total binary
fraction of $\sim 0.55$ is probably a safe lower limit (cf. the
mass-ratio distributions found by Duquennoy \& Mayor 1991; Kouwenhoven
et al. 2007). {\em Depending on exactly which assumptions are made for
the form of the mass-ratio distribution, the total binary fraction
ranges from 0.55 to (more probably) unity.}

\section{Discussion and conclusions}

The CMD of NGC 1818, obtained from {\sl HST} photometry, shows a
clearly asymmetric broadening of the main sequence, which implies that
this cluster contains a large fraction of binary systems. Using the
artificial-star-test method, we estimate that the binary fraction in
the mass range from 1.3 to $1.6 M_\odot$ is $f_{\rm b} \sim 0.35$ for
systems with an approximately flat mass-ratio distribution for
$q>0.4$. This is consistent with a {\em total} binary fraction of F
stars of 0.55 to unity. Elson et al. (1998) found the fraction of
roughly equal-mass ($q \sim 1$) systems in NGC 1818 to be 30--40\% in
the core and 15--25\% outside the core, which is consistent with our
result.

Observationally determined binary fractions are rather scant in
general. Binary fractions are a clear function of stellar type, age,
and environment. We note that our result is quite close to the
fraction of binary dwarfs in the field of the same spectral type,
which is a little smaller than the fraction in Sco OB2 (Kouwenhoven
2006) and much higher than in Galactic globular clusters (e.g., RB97;
Bellazzini et al. 2002). The most recently published relevant results
include the determination by Sana et al. (2008) of the binary fraction
of O-type stars in NGC 6231 (at least 63\%), while Reipurth et
al. (2007) reported a visual-binary fraction of late-type stars in the
Orion Nebula Cluster of only 8.8\%.

At 15--25~Myr old, NGC 1818 is several crossing times old, and the
binary population would be expected to have been modified by dynamical
interactions (see Goodwin et al. 2007; and references therein). In
particular, soft (i.e., wide) binaries would be expected to have been
destroyed by this age. Therefore, the high binary fraction found for F
stars suggests that these binaries are relatively `hard' and able to
survive dynamical encounters.\footnote{We argue that the binaries
in NGC 1818 must be `hard,' since the cluster is dynamically old and
soft binaries are destroyed in roughly a crossing time (e.g., Parker
et al. 2009), so soft binaries cannot be a major component of the
binary population (some may form by capture but will be quickly
destroyed).}  The relatively flat mass-ratio distribution in NGC~1818
compared to similar-mass stars in the loose association Sco OB2
(Kouwenhoven et al. 2007; $\sim q^{-0.4}$) may be evidence for a
difference in the initial populations (see also Sana et al. 2008 for
an alternative mass-ratio distribution biased toward unity). However,
it is more likely to be a product of the different dynamical evolution
of the two populations. The larger number of encounters suffered by
the binary population in NGC~1818 would be expected to disrupt
less-bound (i.e., wide and/or low-$q$) systems, and to form more
equal-mass systems, leading to a mass-ratio distribution more biased
to high $q$.

We conclude that the binary fraction of F stars in the young,
low-metallicity LMC cluster NGC~1818 is high and consistent with the
field and lower-density clusters. This suggests that, at least among
intermediate-mass stars, metallicity down to $\mbox{[Fe/H]} \sim -0.4$
dex does not suppress fragmentation and binary formation, and the
binarity of these stars is at least as high as at solar metallicity.

\acknowledgments

We gratefully acknowledge joint networking funding from the Royal
Society in the UK and the National Natural Science Foundation of China
(NSFC), supporting a UK-China International Joint Project between the
University of Sheffield and the National Astronomical Observatories
(Chinese Academy of Sciences) of China. YH, LD, and QL acknowledge
financial support from NSFC grants 10973015 and 10778719, and the
Ministry of Science and Technology of China grant 2007CB815406. RdG
acknowledges NSFC grant 11043006. He also acknowedges financial
assistance from the Kavli Institute for Astronomy and Astrophysics
prior to moving from the UK to China. This research has made use of
NASA's Astrophysics Data System Abstract Service. We also thank Thijs
Kouwenhoven for useful discussions and valuable input. Finally, RdG
and LD pay tribute to the 2003 Guillermo Haro workshop for initiating
their collaboration, having thus far resulted in 2 PhD theses and
numerous secondary benefits. This turned out to be a career-defining
moment for RdG.

\clearpage
\begin{table}
\caption{Fundamental parameters of NGC 1818}
\begin{center}
\begin{tabular}{lcc}
\hline
 Parameter & Value & Ref. \\
 \hline
 log(age/yr)  & $7.2\pm0.1$ & 2 \\
              & $7.4-7.6$   & 5 \\
 $\mbox{[Fe/H]}$ (dex) & $-0.4$ & 6 \\
 $E(B-V)$ (mag) & 0.05      & 4 \\
 $(m-M)_{0}$ (mag) & 18.58  & 1 \\
 $\log(t_{\rm rh} / {\rm yr})$ & $9.0-9.7$ & 7 \\
 Mass ($M_{\odot}$) & $3\times10^4$ & 3\\
  $R_{\rm core}$ (pc) & 2.56 & 2\\
 $R_{\rm hl}$ (pc) & 2.6 & 2\\
\hline
\end{tabular}
\end{center}
{\sc References:} 1, Castro et al. (2001); 2, de Grijs et al. (2002a);
3, Elson et al. (1987); 4, Hunter et al. (1997); 5, Johnson et
al. (2001); 6, Korn et al. (2000); 7, Santiago et al. (2001).
\end{table}

\clearpage
\begin{figure*}
\includegraphics[width=\columnwidth]{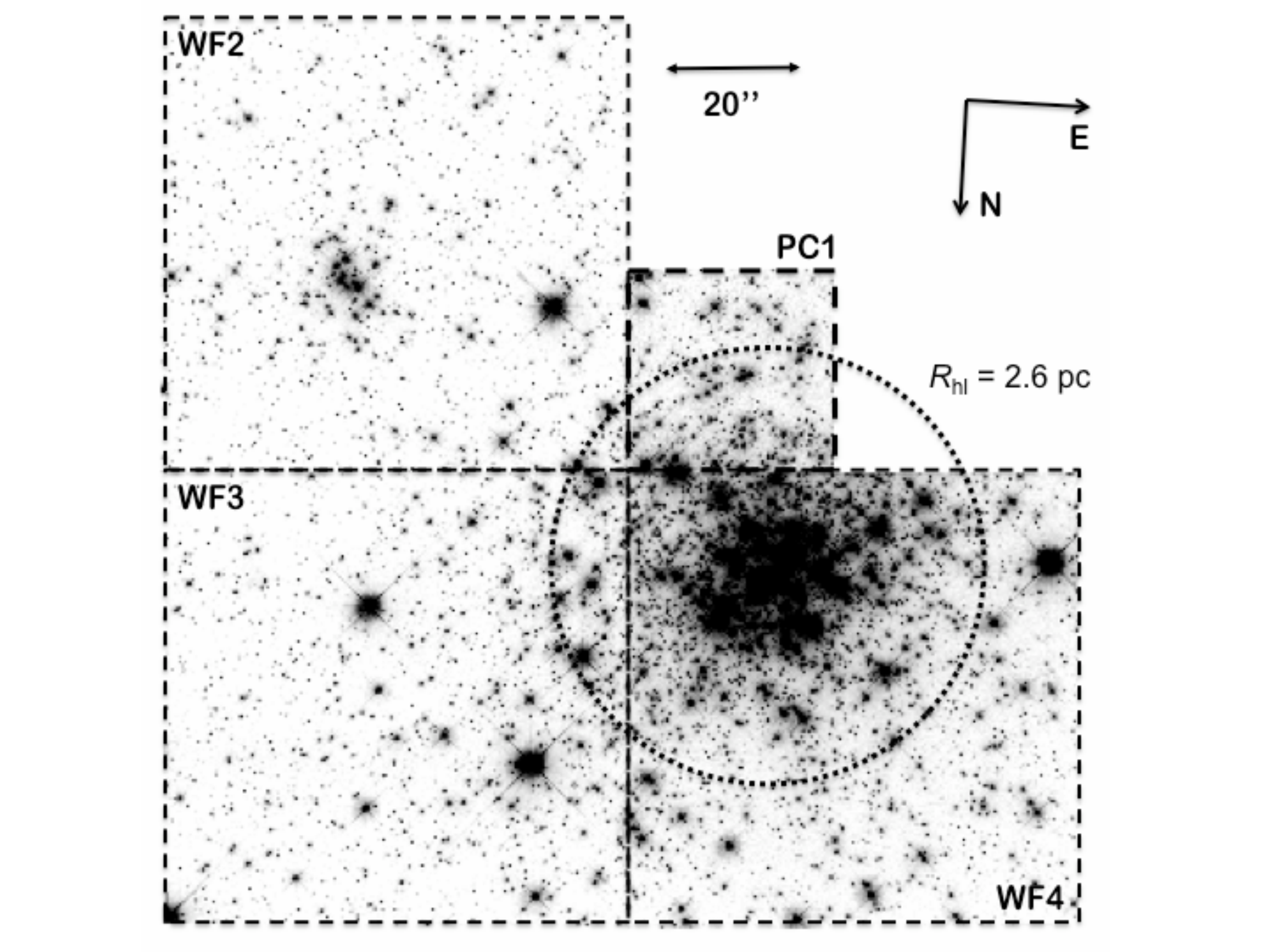}
\caption{Overview of the NGC 1818 field used in this paper. The
individual {\sl HST}/WFPC2 chips are labeled, as is the cluster's
half-light radius (dotted circle).}
\end{figure*}

\begin{figure*}
\includegraphics[width=\columnwidth]{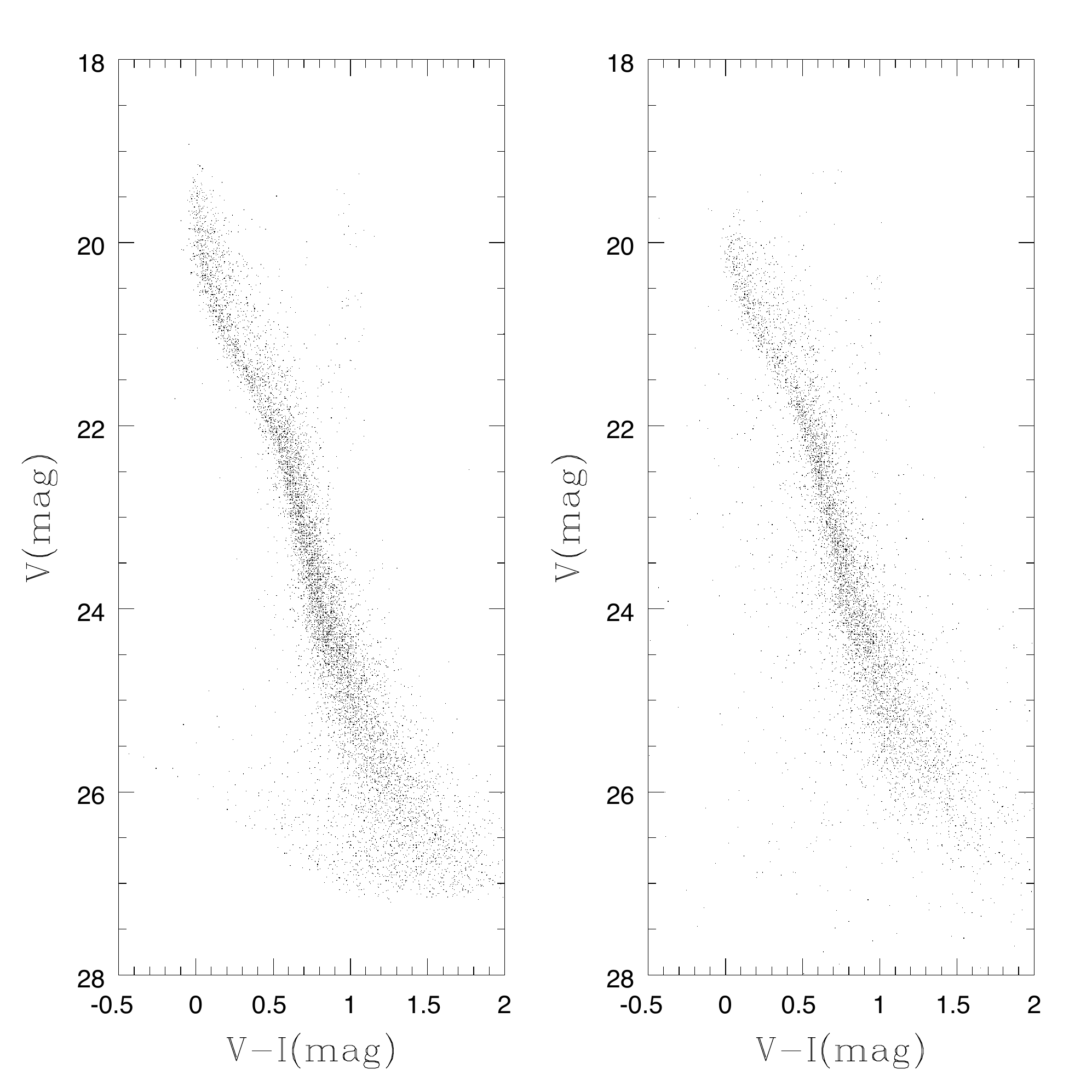}
\caption{{\it (left)} Our newly determined CMD of NGC 1818 at its
half-mass radius. {\it (right)} CMD from Liu et
al. (2009).}
\end{figure*}

\begin{figure*}
\vspace{-9cm}
\begin{center}
\hspace{-1.5cm}
\includegraphics[width=\columnwidth]{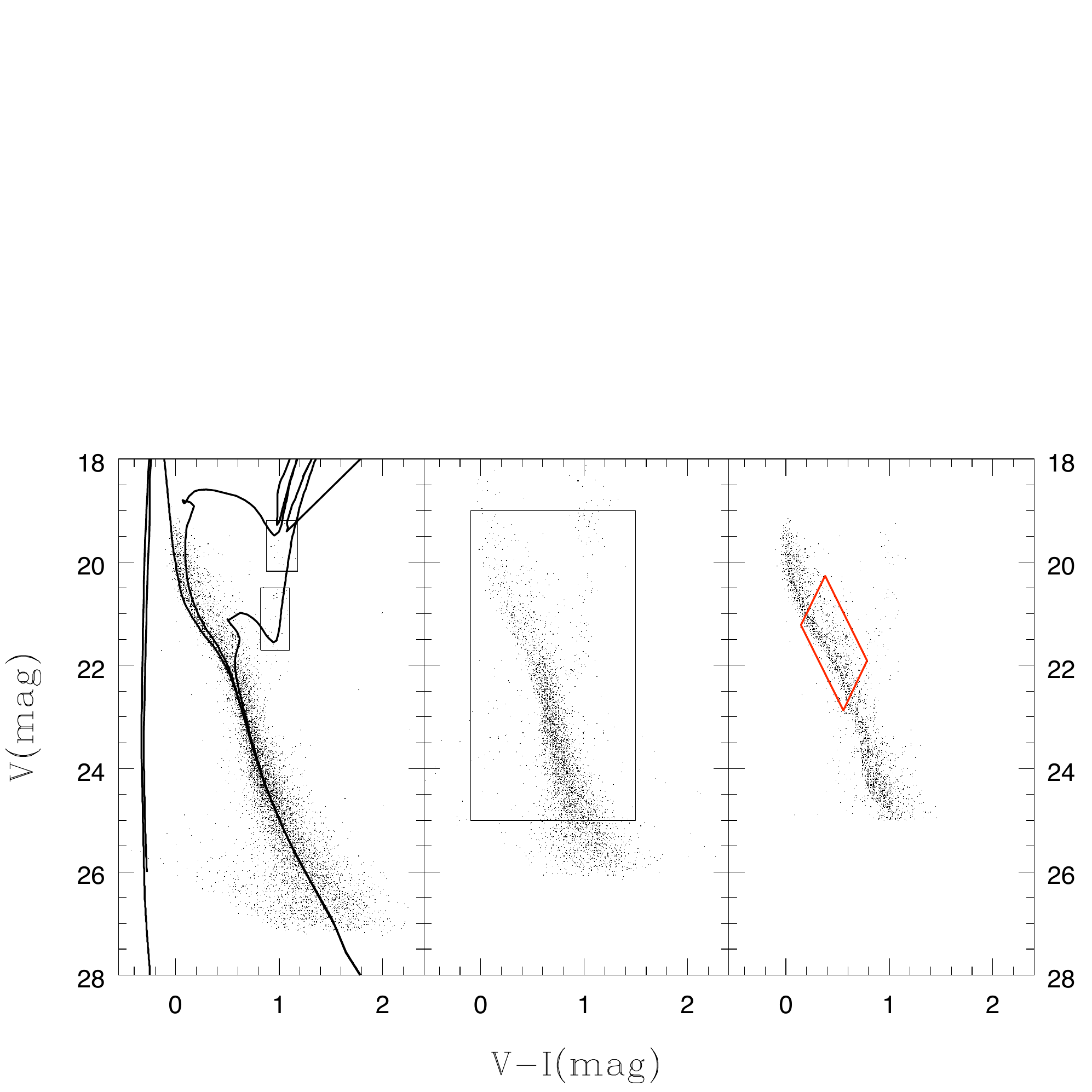}
\end{center}
\vspace{-1.5cm}
\caption{CMDs of NGC~1818 and the background field. The left-hand
panel is the original CMD. A 25~Myr-old isochrone is used to fit the
cluster, and two much older, 0.6~Gyr and 2.5~Gyr isochrones to fit the
background red-clump and red-giant stars. All isochrones shown are for
a metallicity of $Z = 0.008$ (cf. $Z_\odot = 0.019$). The contribution
of the background (field) stars at these evolutionary stages is
estimated from star counts in the two rectangular boxes shown in this
panel (the upper and lower boxes are used for the red-clump and
red-giant stars, respectively). The middle panel shows the CMD of the
background field (from which we only use the stars inside the box
shown in this panel for the field-star correction done in this paper),
and the right-hand panel is the decontaminated CMD of NGC~1818. Our
analysis of the binary fraction of NGC~1818 is confined to the stars
in the parallelogram indicated in the right-hand panel, which covers
stars with masses from 1.3 to $1.6~M_\odot$. We note that in the
region of CMD space of interest (the parallellogram in the right-hand
panel), the completeness fraction of our data is well in excess of
80\% for the faintest stars, irrespective of position in the cluster.}
\end{figure*}

\clearpage
\begin{figure}
\begin{center}
\includegraphics[width=\columnwidth]{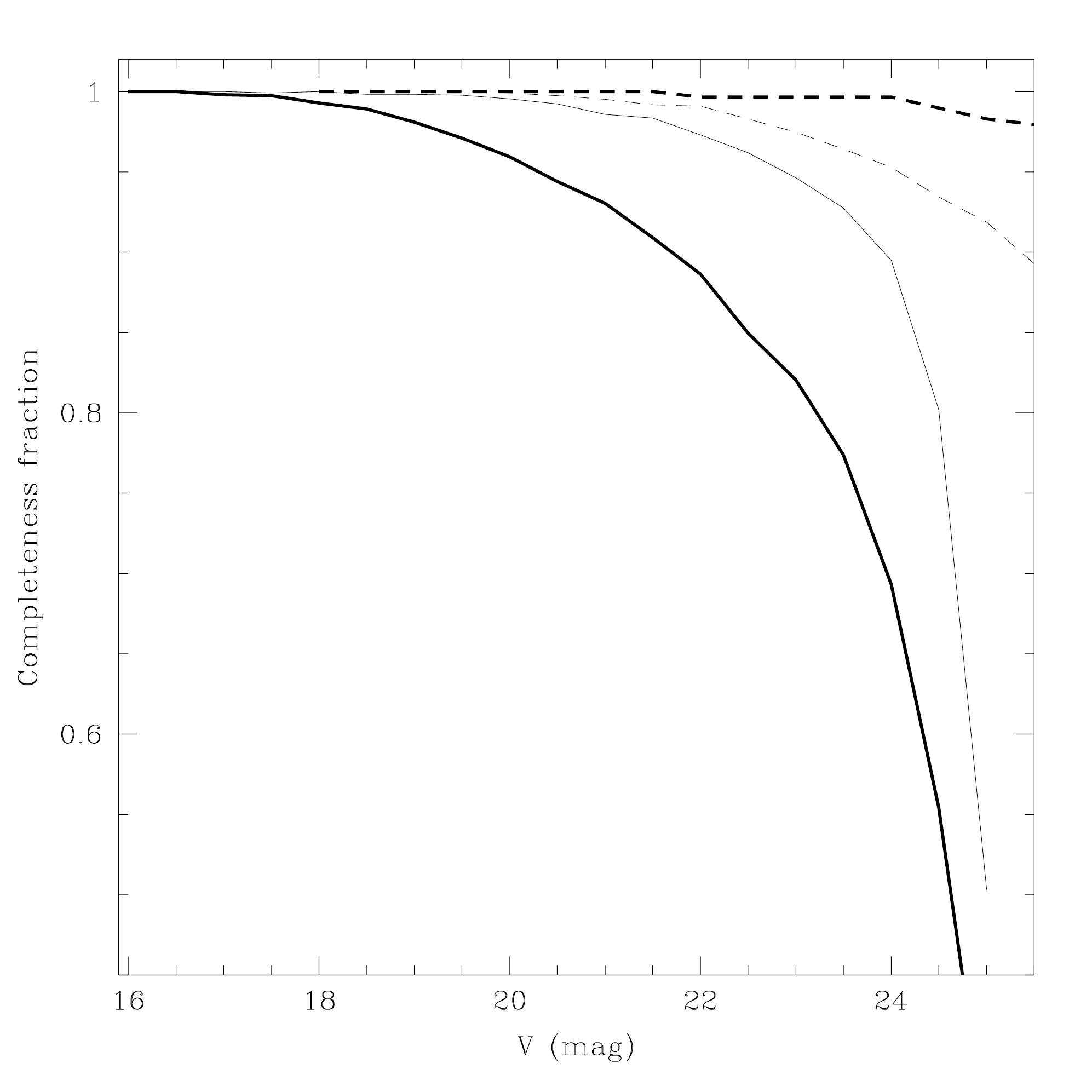}
\end{center}
\caption{Completeness curves for our NGC 1818 observations (solid
lines) and the background field (dashed lines). The thick curves are
for the PC chip and the thin curves represent the data for the WF3
chip.}
\end{figure}

\clearpage
\begin{figure}
\includegraphics[width=\columnwidth]{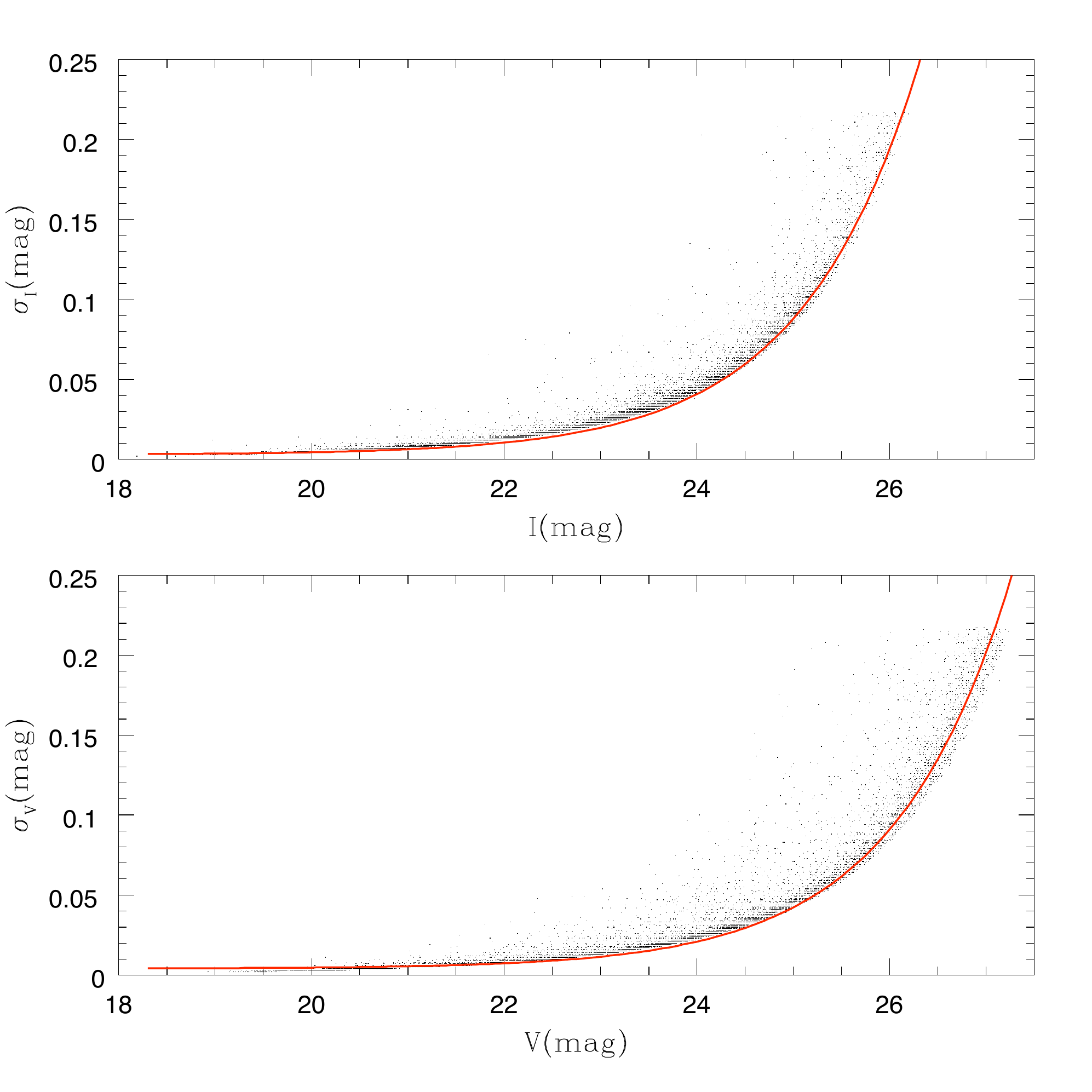}
\caption{Relations between the standard deviations of the photometric
uncertainties and stellar magnitudes for the WF3 chip. The solid lines
are the best exponential fits to the lower boundaries of the data
points for $V, I < 25$ mag.}\label{fig:fig3}
\end{figure}

\begin{figure}
\includegraphics[width=\columnwidth]{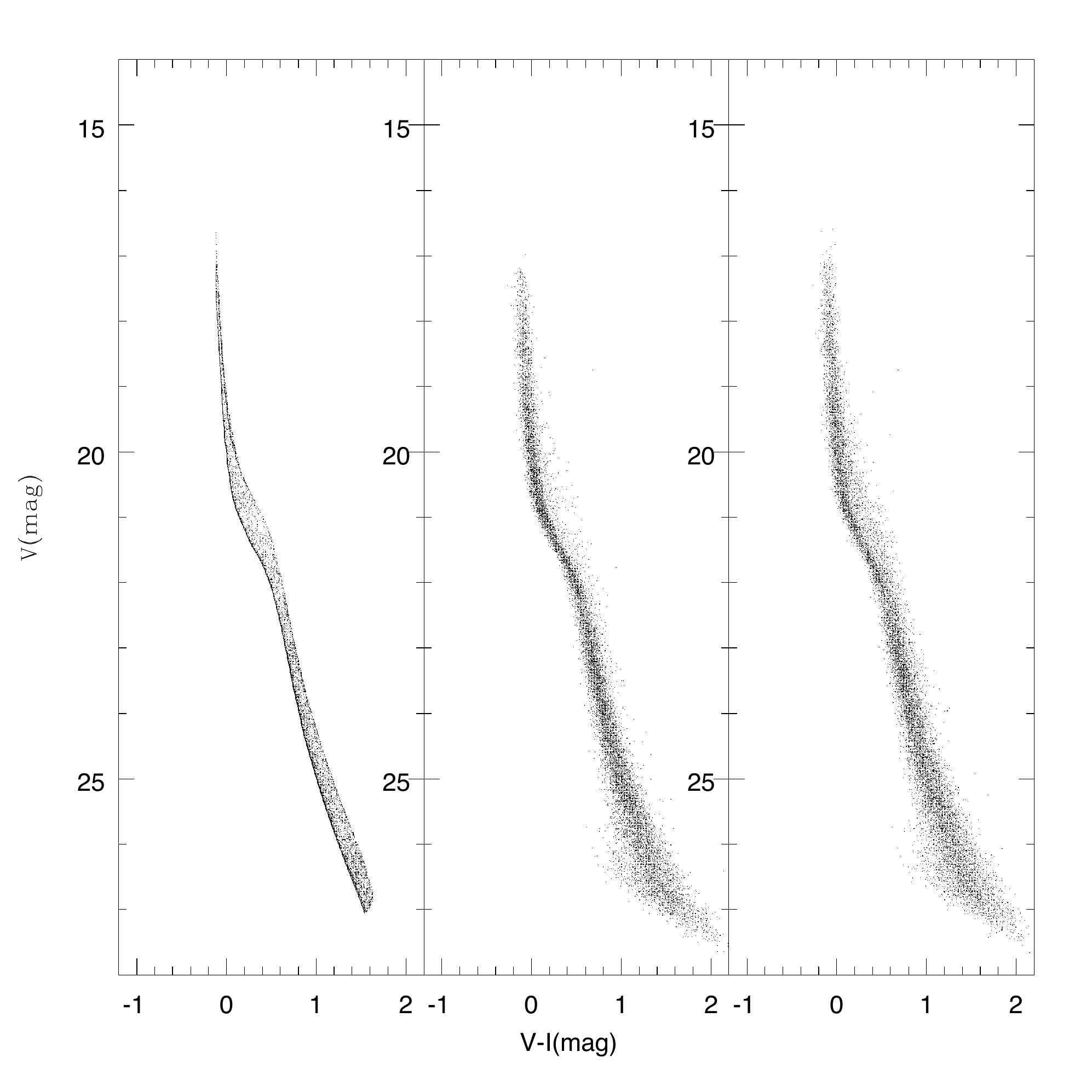}
\caption{{\it (left)} CMD of the artificial stars without the
inclusion of errors, and a 50\% binary fraction. An equal-mass binary
sequence 0.752~mag brighter than the main sequence can be seen
clearly. {\it (middle)} CMD of the artificial stars with Gaussian
photometric errors, but without any binaries. {\it (right)} CMD of the
artificial stars with a 50\% binary fraction {\em and} Gaussian
photometric errors. (Note that without any binaries, a `binary
sequence' is also observed in the CMD. The number of real stars above
the binary sequence plus the $1\sigma$ observational uncertainty is
12\%. For artificial star clusters with a 55\% binary fraction this
number is approximately 7\%, while there are nearly no stars at these
loci for artificial clusters with a 0\% binary
fraction.}
\end{figure}

\begin{figure}
\includegraphics[width=\columnwidth]{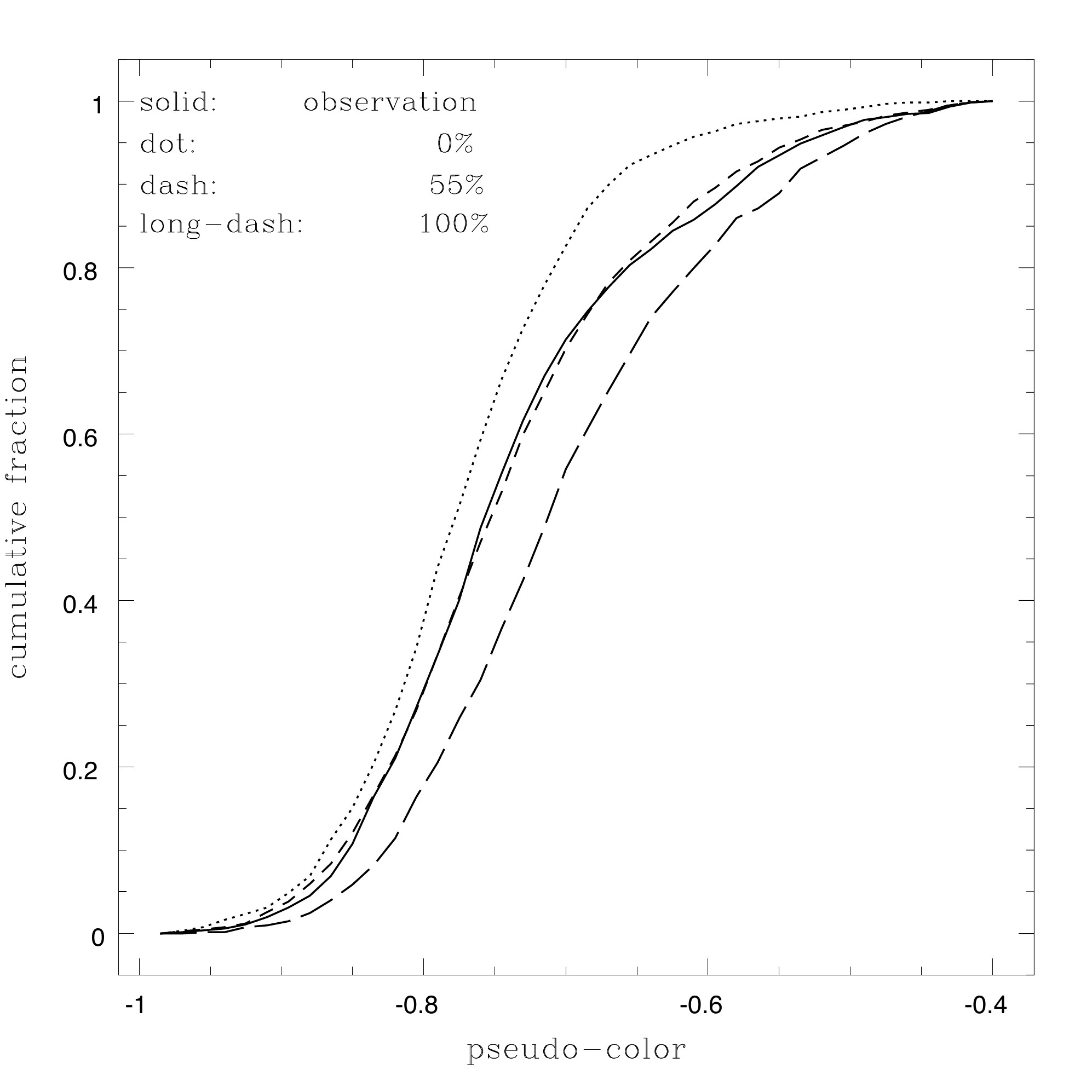}
\caption{Comparison of the observed and best-matching artificial-star
CMDs. {\it (top left)} Background-subtracted observed CMD (equivalent
to the right-hand panel of Fig. 3). {\it (top right)}
Artificial CMD with a 50\% binary fraction (see the right-hand panel
of Fig. 6). The bottom panels show the
pseudo-color--pseudo-magnitude diagrams (see \S 4 for details) of both
the real and artificial stars within the parallellogram region
indicated in Fig. 3.}
\end{figure}

\begin{figure}
\includegraphics[width=\columnwidth]{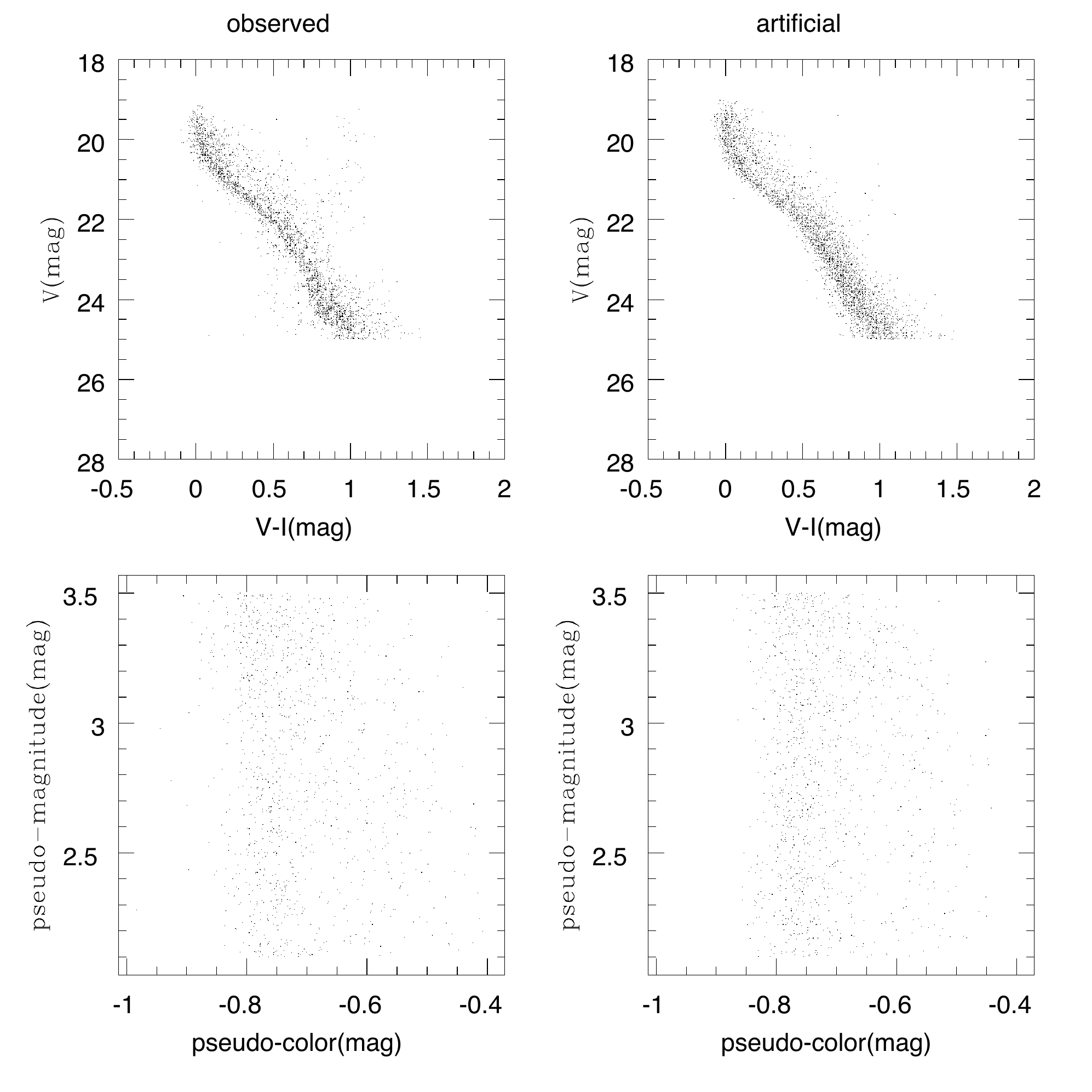}
\caption{Observed cumulative distribution function with pseudo-color
(solid line) compared with an artificial stellar population with zero
and 100\% binary fractions (dotted and long-dashed lines,
respectively), and the best-fitting uniform mass-ratio distribution of
$f_{\rm b} = 0.55 \pm 0.05$ (dashed line).}
\end{figure}

\begin{figure*}
\includegraphics[width=\columnwidth]{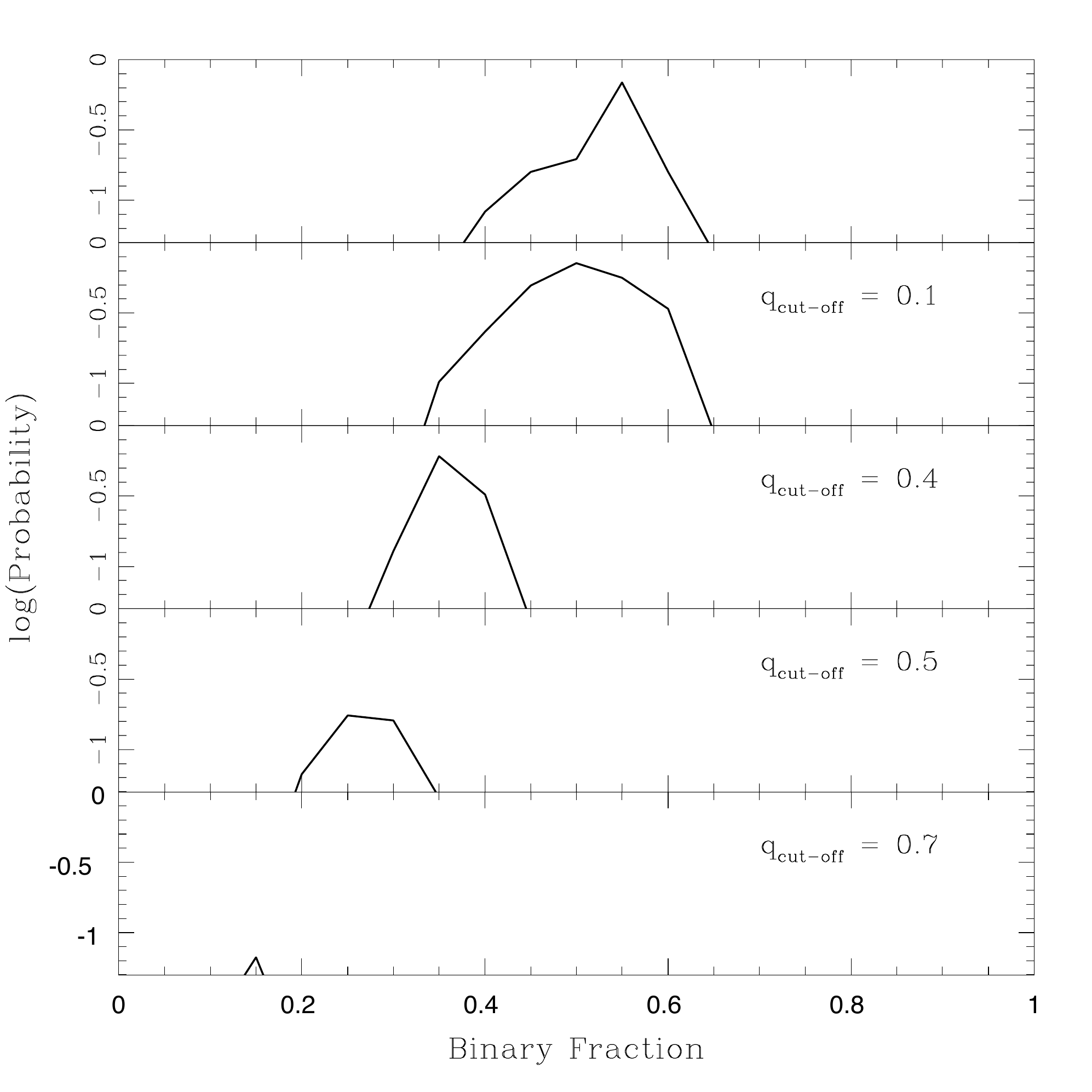}
\caption{Binary fraction and $\chi^2$ probability for different
$q_{\rm cut-off}$. Only probabilities greater than 0.05 are
shown.}
\end{figure*}

\end{document}